\documentclass[10pt, preprint]{emulateapj}		%change back to {aastex} upon submission
\usepackage{color}
\usepackage{style_wei_symbols}
\usepackage{style_wei}

\usepackage{natbib}
\begin{document}

\title{First {\it SDO}/AIA Observation of Solar Prominence Formation Following an Eruption: 
Magnetic Dips and Sustained Condensation and Drainage}

% Magnetic Dips and 
%{First {\it SDO}/AIA Observations of Coronal Condensation Leading to Prominence Formation and Coronal Rain Underneath Nearby Loop Dip}

\author{Wei Liu\altaffilmark{1}\altaffilmark{2}, Thomas E.~Berger\altaffilmark{1},
   and B.~C. Low\altaffilmark{3}}	%, and Roberto Casini\altaffilmark{3}}

\altaffiltext{1}{Lockheed Martin Solar and Astrophysics Laboratory, 	%Department ADBS, 
  Building 252, 3251 Hanover Street, Palo Alto, CA 94304}
\altaffiltext{2}{W.~W.~Hansen Experimental Physics Laboratory, Stanford University, Stanford, CA 94305}
\altaffiltext{3}{High Altitude Observatory, National Center for Atmospheric Research, P.O. Box 3000, Boulder, CO 80307}

\shorttitle{First {\it SDO}/AIA Observations of Coronal Condensation}
\shortauthors{Liu et al.}
\slugcomment{ApJ Letters, Vol. 745, January 2012, in press}

\begin{abstract}	% 250 (300) words limit by ApJL (ApJ)

	%--- niche ---
Imaging solar coronal condensation forming prominences was difficult in the past, 
a situation recently changed by {\it Hinode} and {\it SDO}.
	%--- dip ---
We present the first example observed with {\it SDO}/AIA,	% at 304~\AA,
in which material gradually cools through multiple EUV channels 
in a transequatorial loop system that confines an earlier eruption. 
Nine hours later, this leads to eventual condensation at the dips of these loops,
	%--- prominence: mass rates ---
forming a moderate-size prominence of $\sim$$10^{14} \, {\rm g}$, 
to be compared to the characteristic $10^{15} \, {\rm g}$ mass of a CME.  
The prominence mass is not static but maintained by condensation 
at a high estimated rate of $10^{10} \, {\rm g} \, {\rm s}^{-1}$ 
against a comparable, sustained drainage through numerous vertical downflow threads, 
such that 96\% of the total condensation ($\sim$$10^{15} \, {\rm g}$) is drained in approximately one day.
The mass condensation and drainage rates temporally {\it correlate} with the total prominence mass.
	%--- prominence: downflow vel, acc ---
The downflow velocity has a narrow Gaussian distribution with a mean of $30 \, {\rm km} \, {\rm s}^{-1}$,
while the downward acceleration distribution has an exponential drop with a mean of $\sim$$1/6 \, g_{\odot}$,
	%These downflows are even slower than the nearby coronal rain,
indicating a significant canceling of gravity, possibly by the Lorentz force.
	%--- conclusion, implication ---
Our observations show that a {\it macroscopic quiescent} prominence is {\it microscopically dynamic}, 
involving the passage of a significant mass through it, 
maintained by a continual mass supply against a comparable mass drainage, 
which bears important implications for CME initiation mechanisms 	%in general.
in which mass unloading is important.

\end{abstract}

\keywords{Sun: activity---Sun: corona---Sun: filaments, prominences}
	%ApJ requires <6 keywords

%--- force pdflatex to use letter size paper, from https://engineering.purdue.edu/~mark/puthesis/faq/margins-pdflatex/ ---------------

\setlength{\pdfpageheight}{\paperheight}
\setlength{\pdfpagewidth}{\paperwidth}
 %--- further change default horizontal/vertical margin ---
 % \hoffset = -0.2truein
 % \voffset = -0.2truein

%======================================================================================
\section{Introduction}
\label{sect_intro}

%--- 2011/04/09, Follow my original intro text sent to Tom (he reworded to merge w/ coronal bubble intro in the final proposal)
%--- ~/proposals/2011/2011-03_HR-SHS/wkspc/backup/03-06/sect_topics.tex
%--- for references see 1) Tom's final draft, 2) Fang Cheng, Chen Pengfei textbook, p.241, Section 5.2 3) Markus textbook

	%---what is prominence ------
Solar prominences 	%(or filaments when seen on the disk) are structures of approximately $100$ time denser and cooler
are over dense and cool material 
in the tenuous, hot corona \citep{		%Zirker.promin-review.1989SoPh..119..341Z, 
Tandberg-HanssenE.prominenceBook.1995nsp..book.....T, MartinSF.prominence-review.1998SoPh..182..107M,
Labrosse.prominence-review.2010SSRv..151..243L, MackayDH.prominence-review.2010SSRv..151..333M}.
	%than the ambient hot and tenuous corona in which they are suspended.
	%--- why do we care ---
Proposed mechanisms for prominence formation	%, according to the physical processes involved, fall into two categories:
include: (1) {\it chromospheric transport}	%chromospheric injection: the dense and cool chromospheric material can be directly carried upwards, for example, by pressure-driven 
by siphon flows \citep{Pikel'Ner.siphon-prominence.1971SoPh...17...44P, Engvold.Jensen.siphon-prominence.1977SoPh...52...37E},
	%An.Bao.Wu.siphon-form-promin-I.1988SoPh..115...81A},
magnetic reconnection 	%at the chromospheric level 
\citep{Litvinenko.Martin.injection-to-promin-by-reconn.1999SoPh..190...45L,
WangYM.jet.nautre.promin.1999ApJ...520L..71W},		%, Okamoto.rise-column.2010ApJ...719..583O
or magnetic flux emergence \citep{HuYQ.LiuW.flux-rope-emerge.2000ApJ...540.1119H, Okamoto.emerg-flux-form-prominence.2009ApJ...697..913O},
	%by \Alfven wave impulses (Wu et al. 1983),  => not really, it's about shearing footpoint motion
	%or by upward magnetic convection caused by photospheric converging flows (Titov et al. 1993). => not really, it's convergence flow, compress, current sheet, condensation.
and (2) {\it coronal condensation} due to a diversity of radiatively-driven thermal instabilities
	% a runaway thermal instability as a result of decreased temperature and/or increased density leading to even greater radiative loss 
\citep[][p.277]{ParkerE.thermal-instability-4-coronal-condensation.1953ApJ...117..431P, PriestER.MHDbook.1982QB539.M23P74...}
occurring in current sheets \citep{Smith.Priest.current-sheet-condens-promin.1977SoPh...53...25S, Titov.bald-patch-prominence-form.1993A&A...276..564T}
and coronal loops 	%with mass and/or heat sources at the footpoints 
\citep{WuST.prominence-format-injection-condense.1990SoPh..125..277W,	
MokY.promin.form.MHD.1990ApJ...359..228M,
Antiochos.Klimchuk.FP-heating-prominence.1991ApJ...378..372A,
ChoeLee.promin.form.shearing.1992SoPh..138..291C,
	%AntiochosS.FP-heating-condense-prominence.1999ApJ...512..985A,
Karpen.B-geometry-promin.2003ApJ...593.1187K, 
Karpen.shear-arcade-form-promin.2005ApJ...635.1319K,
Luna-Bennasar.Karpen.2012ApJ.promin.thread.HD}.
	%Karpen.Antiochos.impuls-heating.condens.2008ApJ...676..658K, Xia.ChenPF.promin-form-HD-condens.2011arXiv1106.0094X
	%and other environments \citep{Hood.Priest.thermal-collaps-highP-long-field-line-form-prominence.1979A&A....77..233H,
	%ChoeLee.promin.form.shearing.1992SoPh..138..291C}.

Prominence formation by condensation was once questioned because a quiet, static corona
is an inadequate mass source	%does not have sufficient mass to supply 
\citep{	%Tandberg-Hanssen.insuffic-coronal-mass.4.promin.1986NASCP2442....5T, 
Saito.Tandberg-Hanssen.cavity-mass-insuff-4-promin.1973SoPh...31..105S}.
This does not apply to the solar corona that is known to be dynamic with mass being continually replenished,
	% from the dense chromosphere, 
say, by footpoint heating driving chromospheric evaporation 	%into coronal loops
\citep{Karpen.Antiochos.impuls-heating.condens.2008ApJ...676..658K, 
Xia.ChenPF.promin-form-HD-condens.2011ApJ...737...27X}, 
by ubiquitous spicules \citep{DePontieu.hot-corona-origin.2011Sci...331...55D},
or by thermal convection involving emerging magnetic bubbles \citep{Berger.hot-bubble.2011Natur.472..197B}.
	%It is now widely accepted that condensation is central to prominences.

	%--- niche ---
Condensation was seen in the so-called ``cloud prominences" 
\citep{Engvold.cloud-promin.1976SoPh...49..283E, Tandberg-HanssenE.prominenceBook.1995nsp..book.....T,
LinYong.cloud-promin.2006SPD....37.0121L, Stenborg.EIT-img-process.condens.2008ApJ...674.1201S}
or ``coronal spiders" \citep{Allen.coronal-spider.promin.1998ASPC..150..290A}
	%at typical heights of 100~Mm 
with material streaming out along curved trajectories. 
It was also observed in cooling loops \citep{Landi.loop-cooling-condense.2009ApJ...695..221L}
and proposed to produce coronal rain \citep{	%Levine.Withbroe.Skylab-loop.1977SoPh...51...83L,		%, Kjeldseth-Moe.cooling-loop-review.CDS.1998SoPh..182...73K,
Schrijver.coronal-rain.2001SoPh..198..325S, Muller.model-condensI.2003A&A...411..605M}.
However, quantitative analysis of prominence condensation has 	%lagged behind,	because of % been largely absent,
been hampered by the high off-limb scattered light of ground-based instruments 
and limited spatial/temporal resolution or field of view (FOV) of previous space instruments.  

	%they could constitute a separate category, different from quiescent prominences 
	%with spines and barbs or active region prominences (S. Martin 2011, private communication)

This situation has changed with the launches of the \hinode 
	%Solar Optical Telescope (0\farcs2 resolution)	%\citep[0\farcs2 resolution;][]{Tsuneta.SOT.2008SoPh..249..167T} 
and {\it Solar Dynamics Observatory} (\sdoA).
\sdoA's Atmospheric Imaging Assembly \citep[AIA;][]{LemenJ.AIA.instrum.2011SolPhy}
offers unprecedented capabilities 	%of imaging coronal condensation
with a full Sun FOV at 12~s cadence and $1 \farcs 5$ resolution.
Here we present a first AIA observation of prominence formation, involving magnetic dips and plasma
condensing there to build up the prominence mass and balance 	%a significant 
sustained drainage. Observations of this kind may provide answers to three long-standing questions: 
\begin{enumerate}	%\setlength{\itemsep}{0.0pt}	% or move to final Discuss as concluding remarks for future work.
 \item
How do condensation and drainage control the mass budget of a prominence?

 \item
What are the favorable plasma and magnetic field conditions at the micro and macroscopic scales of a prominence?

 \item
What is the relationship between prominences and other solar activities, such as CMEs and flares?

\end{enumerate}

\begin{figure*}[thbp]      %[thbp]---------------------------
	% \centering
 \epsscale{1.1}
 \plotone{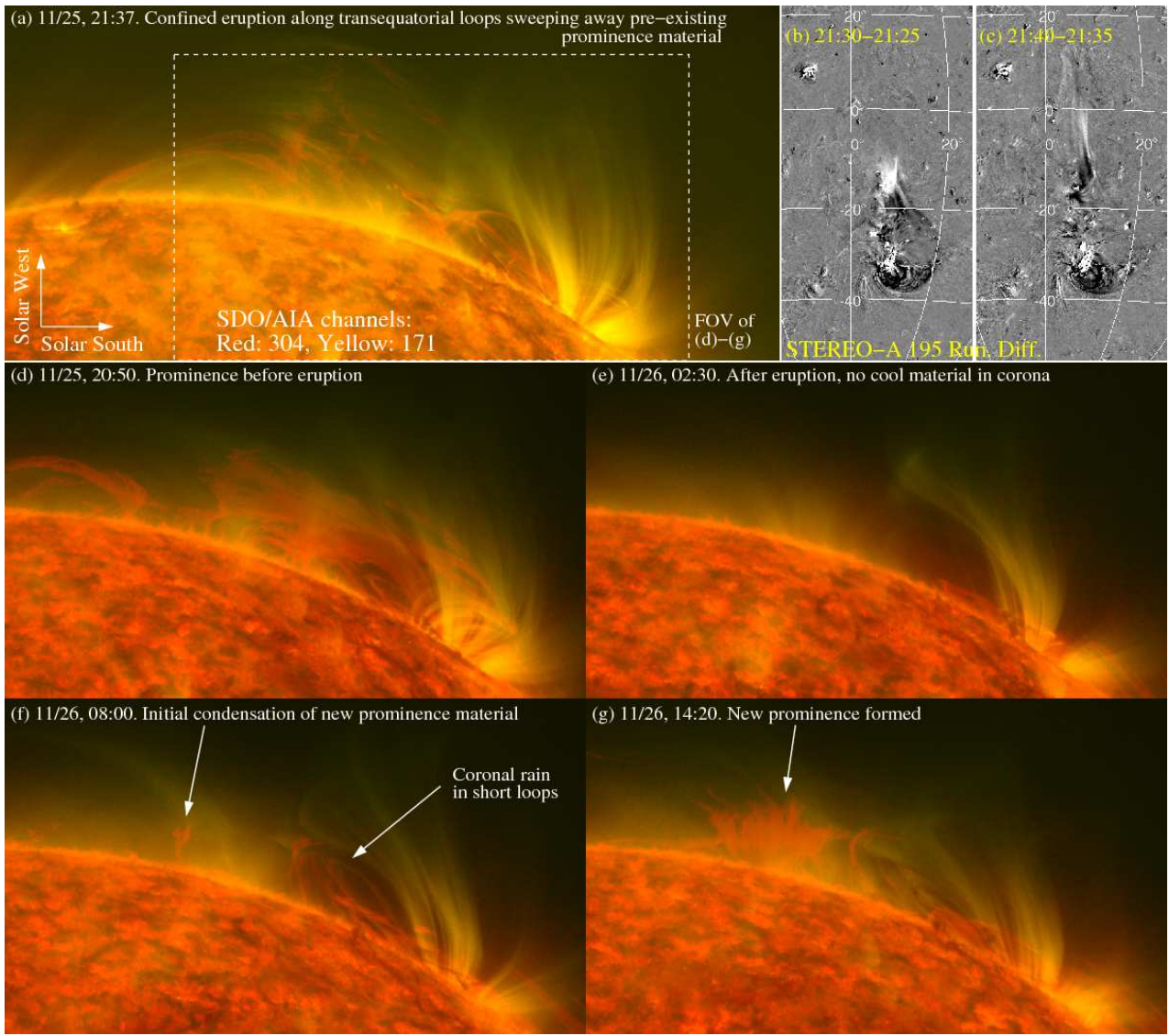}	%{xfig_aia_stereo.eps}
 \caption[]{	%\footnotesize
  (a) Composite AIA 304 (red) and 171 \AA\ (yellow) image, 	% covering the west limb,
 rotated 	%by $90 \degree$ clockwise 
 to Solar West up, showing a confined eruption
 %that sweeps away the pre-existing prominence material 	%and is channeled by transequatorial loops to land on the other hemisphere 
 (Animation~1(A)).
	% This image is rotated 	%by $90 \degree$ clockwise  to Solar West up.
  (b) and (c) 195~\AA\ running difference images from \stereoA~A, $85\degree$ ahead of the Earth, 
 observing the eruption from a top view (Animation~1(B)).
 Note the bulging coronal volume  	%in the south and the trajectory (bright followed by dark)
 and the ejected material 	%near the landing site in the north, 
 forming a rubber syringe shape.
 (d)--(g) Same as (a) but for a smaller FOV (boxed region) 
 showing the event sequence	% before and after the eruption.
 (Animations~1(C) and 1(D)).	% show composite 304/171 \AA\ images and 171 \AA\ images only, respectively.
 } \label{overview.eps}
 \end{figure*}
%

%=======================================================================================
\section{Observational Overview: Prominence Formation After Confined Eruption}		%obs overview, confined eruption
\label{sect_obs}

On 2010/11/25 at about 21:10~UT, a confined eruption is observed by 	%prominence eruption 
\sdoA/AIA and \stereo A in NOAA active region (AR)~11126 (Figures~\ref{overview.eps}(a)--(c); Animations~1(A) and 1(B)).
Most of the pre-existing prominence material is swept away along transequatorial loops
at $\gtrsim$$130 \kmps$ and some lands on the other hemisphere.

At 2010/11/26 05:00~UT, the first sign of new, cool 304~\AA\ material is observed at
the dips in the active region loops and consequently slides downward along underlying loops 
as coronal rain (Figure~\ref{overview.eps}(f); Animations~1(C)).
Two hours later, a nearby condensation at greater rates 
	%adjacent to those elongated transequatorial loops 
leads to the formation of a new prominence (Figures~\ref{overview.eps}(f) and (g)).
	%which is likely the 304~\AA\ counterpart of a so-called ``cloud prominence" 
	%seen in \Ha (S. Martin, private communication). 	%add in ApJ revision, after referee is chosen?
%In spite of the numerous compact downflow streams across the entire prominence, it continues to grow in size. 
    %reaching 100~Mm long and 60~Mm tall.	%reaching a projected length of 100~Mm and height of 60~Mm.
	%Meanwhile, dynamic downflows are observed across the entire prominence.	% bringing mass back to the chromosphere. 
After another day, the prominence gradually shrinks and eventually 
disappears behind the limb.	% due to the solar rotation.

In this Letter, we focus on the condensation and mass drainage in the new
prominence, leaving other interesting aspects of this event for future investigation. 

% Why the eruption is confined, despite the bulging coronal volume
% \citep{FanY.Gibson.loss-confinement.2007ApJ...668.1232F},
% and what caused the condensation at the AR loop dip 
% and subsequent coronal rain are subjects of future investigation.

	% arcsec=arcsec2km('2010/11/26 07:01:38'); 1 arcsec =        715.80750 km, on 2010/11/26 07:01:38
	% help, arcsec* (24.*6)/1e3 = 103.07628; help, arcsec* (80.)/1e3 = 57.264600

%=======================================================================================
\section{Initial Condensation at Magnetic Dips in Transequatorial Loops}
\label{sect_dip}

\begin{figure*}[thbp]      %[thbp]---------------------------
 \epsscale{1.1}
 \plotone{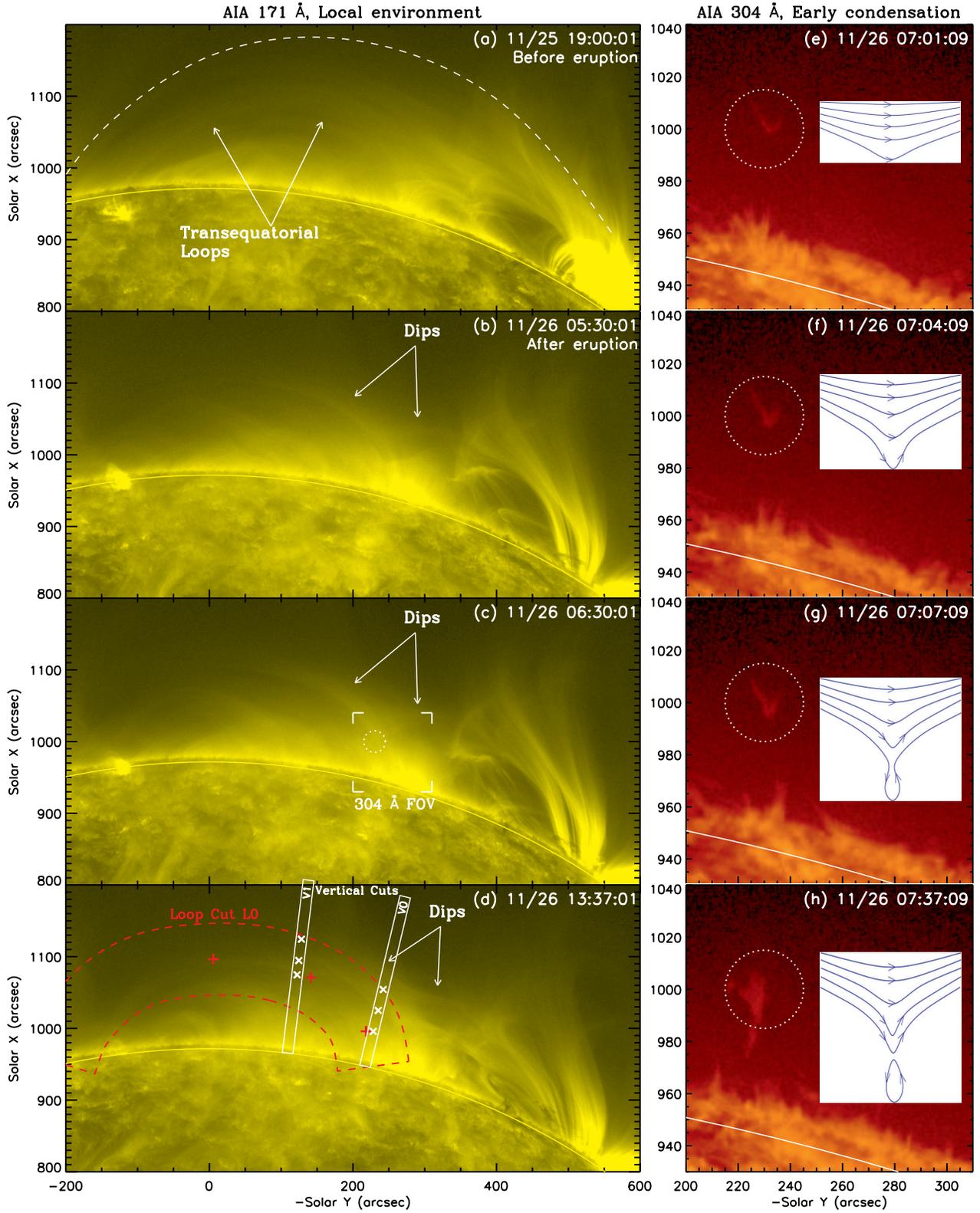}
 \caption[]{
  {\it Left}: AIA 171 \AA\ images showing dips among elongated transequatorial loops.
	% and a magnetic dip.		% at the location of the new prominence.
 Superimposed in (d) are cuts used to obtain space-time diagrams in Figure~\ref{cooling.eps}.
  {\it Right}: Enlarged AIA 304 \AA\ images showing
 the ``V"-shaped initial condensation at the lowest dip.	% indicated by the circle.	% (repeated in (c)). 
 The insets sketch a candidate magnetic configuration.	% suggested by the observations.
 } \label{promin-dip.eps}
 \end{figure*}
Let us first examine the magnetic environment.		% of the prominence condensation.
Prior to the eruption, there are transequatorial loops of various lengths,
best seen at 171~\AA\ in Figure~\ref{promin-dip.eps}(a) and Animation 1(A).
The limb location of the event does not allow accurate determination of the 
magnetic configuration, but potential field extrapolations%
 \footnote{http://sohowww.nascom.nasa.gov/sdb/packages/pfss/l1q$\_$synop}
up to a week earlier
indicate the persistence of such loops that connect the positive trailing polarity of AR~11126
in the south and the diffuse negative polarity in the north.%	 as seen in magnetograms.%
\footnote{see magnetograms on 11/18 at http://www.solarmonitor.org}

	%http://sohowww.nascom.nasa.gov/sdb/packages/pfss/l1q_synop/pfss_20101118_1300_Bfield_20101118_120400_full.png
	%http://www.solarmonitor.org/full_disk.php?date=20101118&type=gong_maglc&indexnum=1

After the eruption, these loops	fade and gradually reappear.
	% become invisible, likely resulting from a combination of density and temperature changes. 
Since 11/26 05:00~UT, one can can identify an apparent depression, likely formed in projection
by shallow dips of numerous loops (Figure~\ref{promin-dip.eps}, left),
where the initial condensation occurs shortly at 07:00~UT.
The cool 304~\AA\ material forms a ``V"-shape (Figure~\ref{promin-dip.eps}, right),
whose lower tip progressively drops and eventually drains down
in a nearly vertical direction, suggesting magnetic reconnection 
(as sketched here) or cross-field slippage (Low et al.~2012b, in preparation).

%=======================================================================================
\section{Post-eruption Cooling in Transequatorial Loops}
\label{sect_cooling}

\begin{figure*}[thbp]      %[thbp]---------------------------
 \epsscale{1.1}
 \plotone{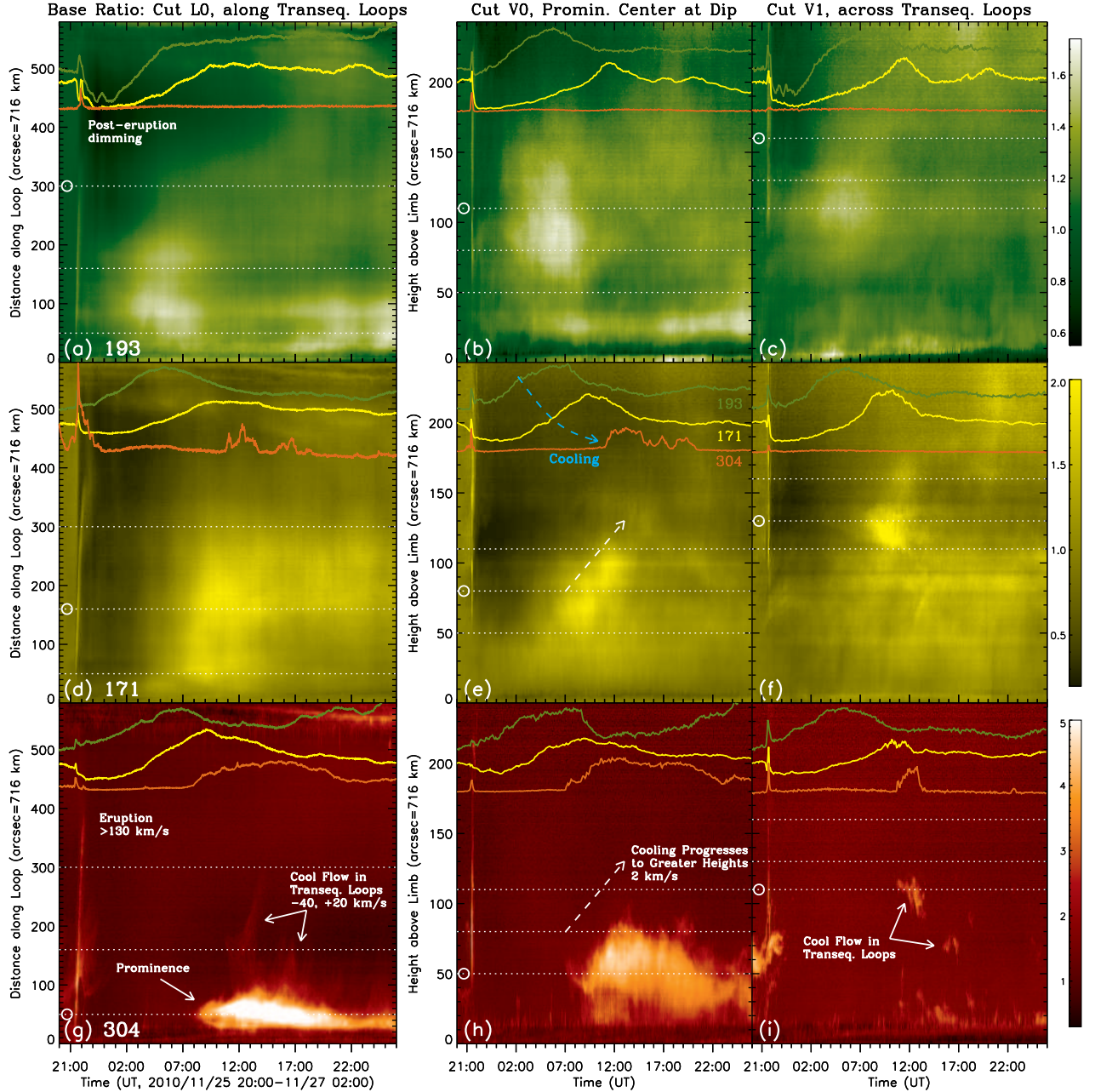}
 \caption[]{
 Base ratio space-time diagrams at 193 (green), 171 (yellow), and 304 \AA\ (red) from 
 the loop shaped cut L0 (left) and vertical cuts V0 (middle) and V1 (right) shown in Figure~\ref{promin-dip.eps}(d).
 Three positions are selected for each cut, as marked with the plus or cross signs there,
 to obtain horizontal slices from the diagram here as indicated by the white-dotted lines.
 The resulting temporal profiles (arbitrarily scaled) in the three channels
 are shown in each panel for the slice marked with an open circle.
 } \label{cooling.eps}
 \end{figure*}
%
%--- space-time technical ---
To track the cooling process and infer the origin of the prominence mass, we placed 
a $100 \arcsec$ ($1 \arcsec = 716 \km$) wide cut L0 along the transequatorial loops to the north
of the dips and two vertical cuts of $15\arcsec$ wide and $243\arcsec$ tall 
from the limb across these loops at the lowest dip (V0) and further north (V1;
Figure~\ref{promin-dip.eps}(d); cuts to the south of the dips yielded similar results).
By averaging image pixels across each cut,	% in 193, 171, and 304~\AA\ images, 	%(hereafter 193, 171, and 304)
we obtained space-time diagrams as shown in Figure~\ref{cooling.eps}
together with selected temporal profiles.
	%, which have peak formation temperatures of $1.6 \E{6}$, $7.9 \E{5}$, and $7.9 \E{4} \K$, respectively

%--- eruption, dimming, brightening, cooling ---%--- 304 prominence as absorption at 193 ---
	%--- brightening ----------
We find a general trend of successive brightening after the eruption
at 193, 171, and then 304~\AA, each delayed by 2--5 hours.
This indicates gradual cooling 	%of material 
across the characteristic temperatures at the response peaks of these broadband channels,	%the characteristic temperatures of these channels, 
from $1.6 \E{6}$ to $7.9 \E{5}$ and $7.9 \E{4} \K$
\citep{Paul.Boerner.initial.AIA.calib.2011SoPh..tmp..316B}.		%\citep{ODwyer.AIA-T-response.2010A&A...521A..21O}. 
	% see /disk2/scr0/weiliu/sdo/docs/AIA/wavelength_T-resp/aia_get_response
At 193~\AA, the initial brightening likely results from mass being
replenished after the eruption and/or cooling from even higher temperatures;	% into this passband.
	%--- 304 prominence as absorption at 193 ---
the later darkening at 9:00~UT and heights $30\arcsec$--$80\arcsec$ 
is primarily due to	%exacerbated by 
absorption by the cool prominence material.

%--- cooling volume ------
Condensation of the 304~\AA\ material occurs only at low altitudes ($<$$100\arcsec$) 
near the dips, while cooling from higher temperatures takes place almost throughout the entire transequatorial loop 
system involving a vast volume (Figure~\ref{cooling.eps}, left).
This implies that these loops likely feed mass to the prominence forming at their dips
that, partly because of the stretched and possibly sheared magnetic field lines
(say, near a polarity inversion line) and thus reduced field-aligned thermal conduction, 
may facilitate further cooling and condensation (Low et al.~2012a, in preparation).
We detected no systematic cooling flow in the loops though,	% during this cooling process,
possibly due to the insensitivity of AIA (passband imager) 
to subtle flows without distinct mass blobs,	%detectable intensity variations
		%(passband, not spectroscopic, imager) 	%at moderate temperatures % steady-state like 
		% which might be due to our limited temperature coverage or the flow is steady-state and thus causing no detectable intensity variations.
	%--- occasional rain: up/downflows, delay in the same loop ---	We found only 
except for occasional 	%(e.g., 11/26 11:00--14:00 and 16:00--17:00~UT) 
brief downflows and upflows of 304~\AA\ material 
	% at typical velocities of $40$ and $20 \kmps$, respectively 
(Figures~\ref{cooling.eps}(g)).
	%, which occur well above the prominence long after its initial condensation,	%(07:00~UT)
	%with negligible contribution to the prominence mass. 
	%agrees with the timescales 
	%of simulated recurrent condensation in loops of similar lengths \citep{Karpen.no-dip-promin.2001ApJ...553L..85K}.
The 9~hr elapsed from the eruption to the initial condensation is on the same order of
magnitude as the condensation timescales in thermal nonequilibrium simulations
\citep{Karpen.shear-arcade-form-promin.2005ApJ...635.1319K}.

  	%--- cooling delay at greater heights, same/different loops ---
Condensation at 304~\AA\ occurs first at a height of $60 \arcsec$
and progresses toward higher loops		% at larger altitudes 		%up to $100 \arcsec$ 
at $2 \kmps$ ($\gg 0.3 \kmps$, the projected speed of the
solar rotation there), with a similar trend at 171~\AA\ (Figures~\ref{cooling.eps}(e) and (h)).
	%This speed is small but real, as it is significantly greater than the 
	%velocity $< 0.3 \kmps$ projected onto the plane of the sky due to 
	%the solar rotation at a central meridian distance $> 80 \degree$ estimated for this event.
	 % see /disk2/scr0/weiliu/sdo/events/04_filament/10-11-26_flmt-form/4paper/dip-time-slice/v0_2p/README
	%--- n^2 dependence ---
We speculate that those low-lying loops condense first because of their shorter lengths 
\citep{Karpen.Antiochos.impuls-heating.condens.2008ApJ...676..658K} and/or
their greater densities from gravitational stratification (as radiative loss $\propto$$n^2$).

	%As time proceeds, loops at greater altitudes sags over the lower-lying loops layer by layer
	%and then condense to prominence temperatures.

%=======================================================================================
\section{Mass Budget of Condensation and Drainage}
\label{sect_mass}

%In what follows, we investigate coronal condensation and mass downflows in the new
%prominence as our focus and at the local dip for comparison.	% individually. 	%with the former being our main purpose.
	%Our main purpose is to infer the amount and rate of mass that is condensed into the prominence.

	% because of their different characteristics. 
	%For the former, our objective is to estimate at what rate mass condensation takes place.

%---------------------------------------------------------------------------------
%\subsection{Condensation in New Prominence}
%\label{subsect_promin}

In this prominence, we find no sign of rising bubbles
\citep{BergerT.promin-plume.2008ApJ...676L..89B, BergerT.bubble-RT-instable.2010ApJ...716.1288B, 
Berger.hot-bubble.2011Natur.472..197B} that can contribute to the prominence mass.
If we further neglect other possible mass sources,
the interplay between the mass condensation rate $\dot{M}_{\rm cond}$ and downflow rate $\dot{M}_{\rm drain} $
would determine the prominence mass change rate 
 \beq
 \dot{M}_{\rm prom} = \dot{M}_{\rm cond} - \dot{M}_{\rm drain}.
 \label{mass.conserve.eq} \eeq
The condensation rate $\dot{M}_{\rm cond}$, which cannot be measured directly,
can thus be inferred by measuring $\dot{M}_{\rm prom}$ and $\dot{M}_{\rm drain} $ independently, as described below
in three steps.

%----------------------------------------------------------
\subsection{Prominence Mass Budget $\dot{M}_{\rm prom}$}
\label{subsect_mass}

This prominence consists of numerous nearly vertical threads of
typically $d_{\rm thrd}=2 \arcsec$ (1.43~Mm) thick (see also \citealt{BergerT.promin-plume.2008ApJ...676L..89B}).
The observed optically thick 304~\AA\ emission originates from a layer known as 
the prominence-coronal transition region (PCTR) surrounding a denser, cooler
\Ha emitting core \cite[e.g.,][]{LinYong.fine-promin-thread.2005SoPh..226..239L, LinYong.promin-thread-review.2011SSRv..158..237L},
later seen in Mauna Loa data (not shown). However, lacking sufficient \Ha coverage of the event, 
we chose to use a typical PCTR electron density $n_{\rm e}= 8 \E{9} \pcmc$
\citep{Labrosse.prominence-review.2010SSRv..151..243L}	%Table 4 geometric mean 
and thus $\rho= n_{\rm e} m_{\rm p} = 1.3 \E{-14} \g\pcmc$
as a lower limit for the entire prominence.
We also conservatively assumed that the observed threads do not overlap each other.

	%Due to lack of \Ha coverage, we chose to use the observed PCTR to obtain a lower limit of the entire prominence.
	%neutral hydrogen number density $	%$n_{\rm H}= 3 \E{10} \pcmc$
	%electron density $n_e= 10^{11} \pcmc$ ($\rho= 1.7 \E{-13} \g\pcmc$),	%Labrosse (2010) Eq. (18) on p. 276

We selected a sector-shaped area to enclose the bulk of the prominence (see Figure~\ref{downflow-promin.eps}(c)).
The sector covers a height range of $25\arcsec$--$110\arcsec$ above the limb	%($1\arcsec=0.716 \Mm$)
and a latitude range of $8.3 \degree$ ($144 \arcsec$ wide on its bottom edge).
We then integrated the area $A_{\rm prom}$ occupied by the prominence material inside the sector for all pixels above a threshold
selected at $3\sigma$ above background.	% at 11/26 02:27~UT when no prominence material is detected. 
We obtained a lower limit to the prominence mass
 %\footnote{
 %Assumption (1) may underestimate of the line-of-sight (LOS) thickness and thus the {\it absolute} total mass.
 %However, since this assumption is also self-consistently taken for the mass drainage rate measurement
 %below, our result about the {\it relative} contribution of downflows and condensation is still valid. 
	 %What we may miss is a scaling factor that accounts for the {\it absolute}  LOS thickness and mass of the prominence.
	 %The \ion{He}{2}~II 304~\AA\ emission is optically thick, and thus we only observe the outermost layer
	 %of the prominence body. What we do not see is not counted in either the mass estimate or the drainage rate calculation below.
	% } \footnote{
 %Assumption (2) takes a conservative value of the density,
 %considering the range of the measured electron density $10^{9}$--$10^{11} \pcmc$ and 
 %low degree of ionization 0.2--0.9 (\citealt{Labrosse.prominence-review.2010SSRv..151..243L}; 
 %even 0.09, \citealt{Landman.0.09-promin-ionization-degree.1984ApJ...279..438L}).
 %}
 \beq 
 M_{\rm prom}= \rho A_{\rm prom} d_{\rm thrd},
 \label{mass-history.eq} \eeq
as shown in Figure~\ref{mass-rates.eps}(a). 
	%The milestones of the event described above, 
	%such as the passage of the eruption,	% and initial condensation at 11/26 07:00~UT, 
	%are evident here. 	%in this figure. 
After the initial condensation,	% at 11/26 07:00~UT, 
the prominence mass continues to increase and reaches the maximum at 16:01~UT,
 $$ % \beq 
 M_{\rm prom, \, max}=6.0 \E{13} \g ,		%2.2 \E{14}	%7.5
 $$ % \label{M_max_eq} \eeq
which is on the same order of magnitude as that determined from	% \sohoA/EIT \ion{Fe}{12} 195~\AA\ 
coronal absorption for a prominence of similar size \citep{GilbertH.promin-mass-195Absorption.2005ApJ...618..524G}. 
	%The mass decrease afterwards results from a reduced condensation rate (see Section~\ref{subsect_condense}).
 	%There is a second episode of rapid condensation at $\sim$22:00~UT, followed by
	%The slow decrease near 11/27 12:00~UT		% of total mass caused by reduced condensation 
	%is mainly due to increasing limb occultation from the solar rotation.	
Taking time derivative gives the mass change rate $\dot{M}_{\rm prom}$ shown in Figure~\ref{mass-rates.eps}(c),
which has a mean of $\langle \dot{M}_{\rm prom} \rangle = -5.6 \E{7} \g \ps$ 		%-2.1 \E{8}%-7.0 \E{8} 
and a maximum at 11:37~UT,  
 $$ %\beq 
 \dot{M}_{\rm prom, \, max}= 7.6 \E{9} \g \ps .	%2.8 \E{10}	%9.5
 $$ %\label{M_dot_max_eq} \eeq
	 %that is 2 orders of magnitude smaller and effectively zero, as expected for the life cycle of the prominence.

	% see data at: /10-11-26_flmt-form/4paper/mass-rate/v0_4Tahoe/*.dat
	%A linear fit to the data in the duration 07:00--14:00~UT gives an average mass gain rate of $\dot{M}= 2.6 \E{10} \g \ps$.

%----------------------------------------------------------
\subsection{Mass drainage Rate $\dot{M}_{\rm drain}$}
\label{subsect_drain}

\begin{figure*}[thbp]      %[thbp]---------------------------
 %\epsscale{1.}
 \epsscale{0.57}
 	%\epsscale{0.55}		%for \begin{figure*} sitting side by side
 \plotone{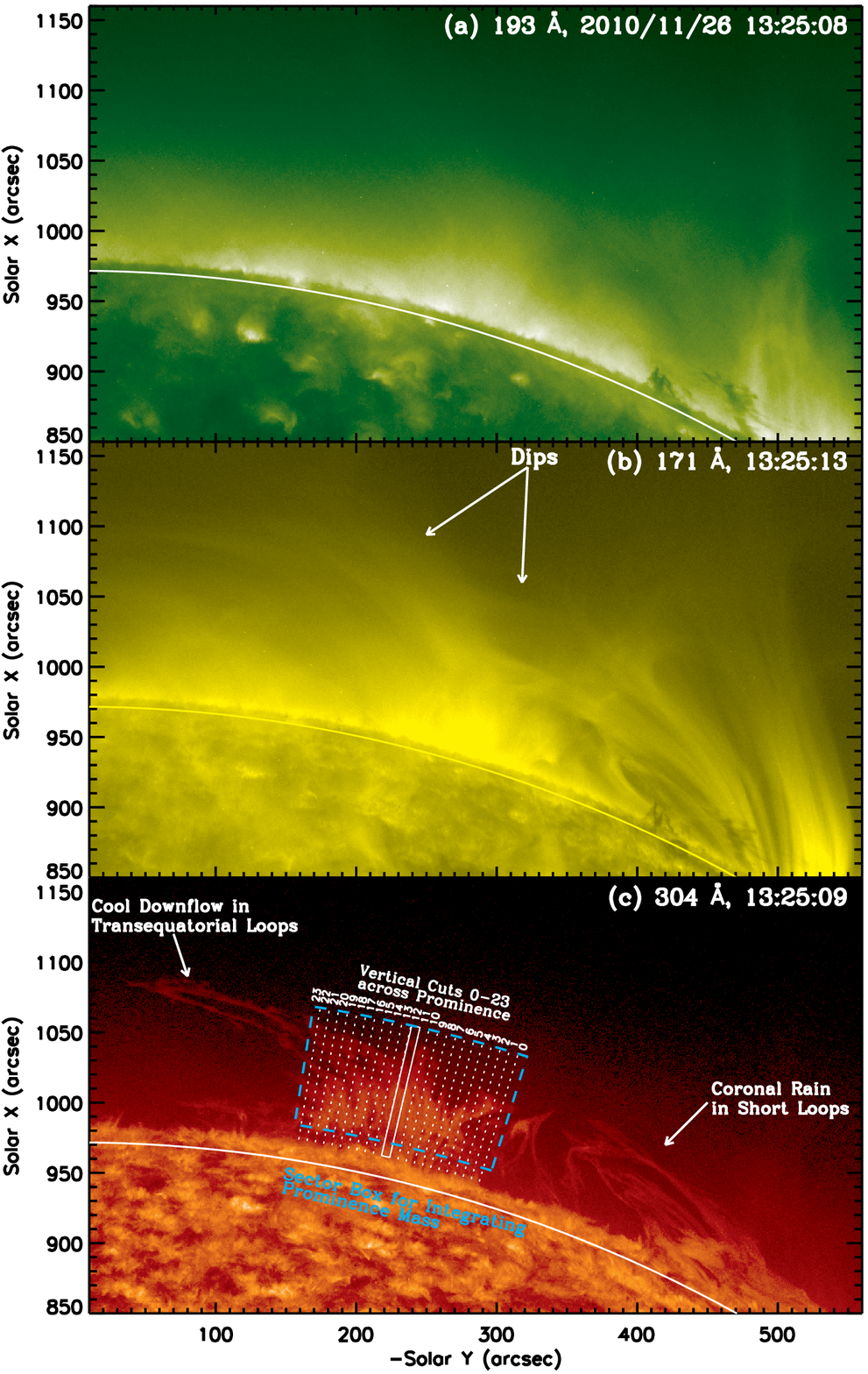}	%{promin_mosaic_color.eps}
 	%\epsscale{0.6}	%for \begin{figure*} sitting side by side
 \plotone{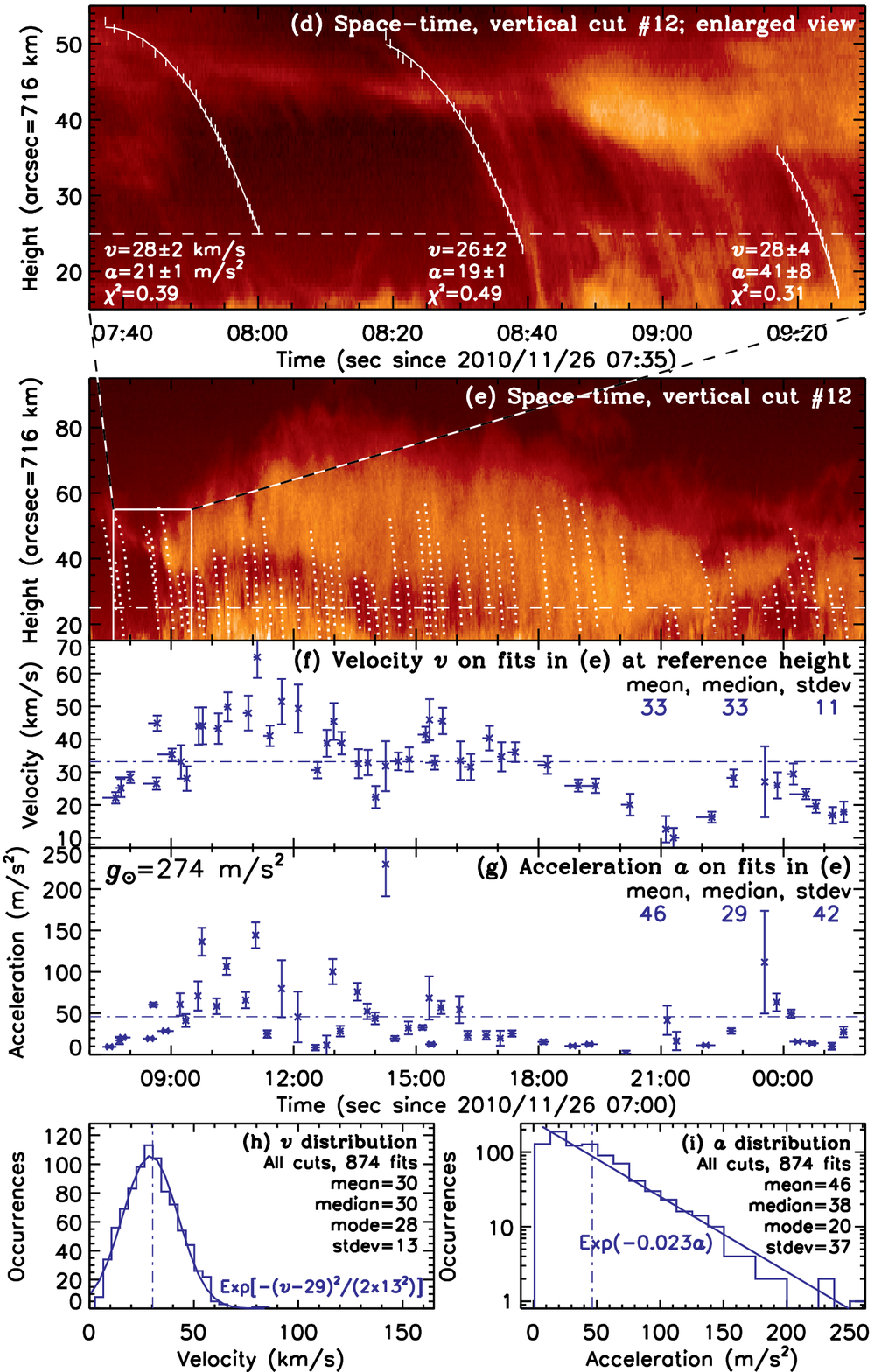}	%{cmpr_time_slice_vel_1126.eps}
 \caption[]{	%\footnotesize
   (a)--(c) Simultaneous AIA 193, 171, and 304 \AA\ images	% (rotated to solar west up) 
 showing the neighborhood of the new prominence.
	 %Overlaid in (c) are a sector-shaped area for integrating the prominence mass
	 %and 24 vertical cuts for measuring the mass downflow rate.	% as shown in Figure~\ref{mass-rates.eps}.
   (d) Enlarged space-time diagram of (e) in the boxed region superimposed with selected fits.
   (e) Space-time diagram obtained from a typical vertical cut (\#12, marked with solid lines in (c);
 see Animation~4 for the diagrams of all vertical cuts).
 White dotted lines are parabolic fits to downflow trajectories. The horizontal dashed
 line marks the reference height $h_{\rm ref} = 25 \arcsec$ of the bottom edge of the sector at which downflow velocities are measured.
   (f) and (g) Velocities and accelerations from the fits in (d) as a function of time. 
 Each short horizontal bar indicates the duration of the fit. 
   (h) and (i) Histograms of velocities and accelerations from 874 measurements obtained from all 24 vertical cuts,
 whose distributions are fitted with a Gaussian and exponential function, respectively.
 The dot-dashed lines in (f)--(i) mark the corresponding mean values.
 } \label{downflow-promin.eps}
 \end{figure*}
%
%--- take vertical cuts to get space-time diagrams ----
To estimate the drainage rate of mass that leaves the sector-shaped area, 
we need to obtain the downflow velocity $v(s, t)$	% across the prominence 
at position $s$ on the bottom edge of the sector and at time $t$,
and then spatially integrate the mass flux crossing this edge. 
Here we assume that the warm PCTR and cool core of the prominence thread 
have the same downflow velocity		%, as suggested by their similar observed values 
\citep{Labrosse.prominence-review.2010SSRv..151..243L}.
To measure $v(s, t)$, we divided the sector into 24 vertical cuts of $6 \arcsec$ wide%
 \footnote{The $6 \arcsec$ width of the vertical cuts was chosen to be as small as possible and yet wide
 enough to enclose most of the prominence threads	% of typically $2 \arcsec$ thick,	%$d_{\rm thrd}=2 \arcsec$, 
 that are close to but not exactly vertical.
 }
that start at $15\arcsec$ above the limb to avoid spicules	%the top of the chromosphere (defined at $15\arcsec$ above the photospheric limb)
and discretize the bottom sector edge into distance bins $s_i$ ($i=0,1,...,23$), 
as shown in Figure~\ref{downflow-promin.eps}(c). 
For each cut, we obtained a space-time diagram 
for a 19~hr duration from 11/26 07:00 to 11/27~02:00~UT.	% at a 12~s cadence.

%--- Cut 12 as an example: space-time diagram and parabolic fits ----
As an example, the space-time diagram (Figures~\ref{downflow-promin.eps}(d) and (e))
of cut \#12 located near the middle of the prominence 
	%together with an enlarged view in Figure~\ref{downflow-promin.eps}(d).
%In general, this diagram can be divided into two regions: an upper portion showing	% relatively stationary 
%dense material in the main body of the prominence and a lower portion showing 
shows numerous, episodic downward curved trajectories.
Each trajectory tracks the height of a blob of mass with time 
as it descends within the vertical cut.	% representing mass downflows with acceleration.	% from the prominence body.
The descent covers a height range of several tens of arcseconds and lasts for a few minutes to half an hour.
	%For a well-defined trajectory, 	%we obtained its space-time positions by mouse clicks,
We fitted each well-defined trajectory with a parabolic function,
as shown in Figure~\ref{downflow-promin.eps}(d),
and determined the average acceleration $a_{\rm prom}$ and
velocity $v_{\rm prom}$ at the reference height $h_{\rm ref} = 25 \arcsec$ (horizontal dashed line) 
of the bottom edge of the sector. 
	%--- could remove this statement about uncertainty later if we need to save space ---
	%We adopted a positional uncertainty at the AIA pixel size of $0\farcs6$,
	%which often gives $\chi ^2$ values smaller than unity, indicating a possible overestimate of the uncertainty. 
%
	%A few example fits are shown in Figure~\ref{downflow-promin.eps}(c).
We exhaustively repeated such fits throughout the space-time diagram, 
only skipping an adjacent trajectory of a similar slope within any 10~minutes interval.
The resulting fits are overlaid in Figure~\ref{downflow-promin.eps}(e),
and the fitted velocities and accelerations are shown in Figures~\ref{downflow-promin.eps}(f) and (g).
	%We find a relatively narrow distribution of the velocity, with a mean (median) of $33 \kmps$
	%and standard deviation of only $11 \kmps$, while the acceleration has a broader distribution
	%with a mean of $46 \m \pss$ and a similar standard deviation.
Finally, we linearly interpolated the fitted velocities in time and obtained 	%a measured velocity 
$v(s_{12}, t)$ covering the full duration.

%--- all cuts statistics: vel and acc from parabolic fits ----
We applied this procedure to all vertical cuts across the prominence and obtained a statistically significant sample
of 874 parabolic downflow trajectories. As shown in Figure~\ref{downflow-promin.eps}(h),
the fitted velocities have a narrow Gaussian distribution, with a mean and median of 
$\langle v_{\rm prom} \rangle= 30 \kmps $, mode of $28 \kmps $, and standard deviation 
of $13 \kmps $.	% $\sigma_{v_{\rm prom}}= 13 \kmps $.
	%It is well fitted with of a Gaussian of similar parameters.
In contrast, the accelerations (Figure~\ref{downflow-promin.eps}(i))
have an exponential distribution $\propto e^{-0.023 a}$, 
with a mean of $\langle a_{\rm prom} \rangle=46 \m \pss = 0.17 g_\Sun \approx 1/6 g_\Sun$, 
a standard deviation of $37 \m \pss$ $(0.14 g_\Sun)$, 
and a small mode of $20 \m \pss$ $(0.07 g_\Sun)$, where $g_\Sun = 274 \m \pss$
is the solar surface gravitational constant.
	%This distribution has a small mode of $20 \m \pss (0.07 g_\Sun) $ and a steep exponential
	%drop $\propto e^{-0.023 a}$.
%
%
\begin{figure}[thbp]      %[thbp]---------------------------
 \epsscale{1.1}
 \plotone{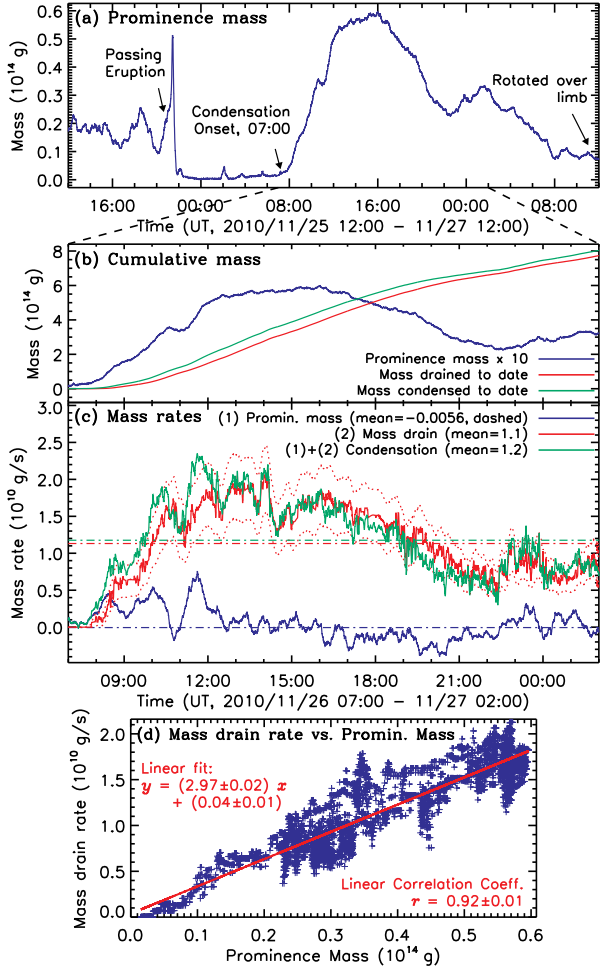}	%{plot_mass_rate.eps}
 \caption[]{	%\footnotesize
   (a) Temporal variation over 2 days of the estimated prominence mass $M_{\rm prom}$.
 	%The sharp spike at 11/25 21:26~UT is due to the passage of prominence material swept by the eruption.
   (b) Prominence mass budget (blue, repeated from (a) but scaled by a factor of $10$)
 and cumulative mass drained (red) and condensed (green) to date.
   (c) Time derivatives of the curves in (b). 		%and the inferred mass condensation rate (green).
 The red dotted lines mark the bounds of uncertainties of the mass drainage rate
 estimated from downflow velocity uncertainties		% in the fitted downflow velocities
 (see, e.g., Figure~\ref{downflow-promin.eps}(f)). The horizontal dot-dashed lines indicate
 the corresponding mean values.
   (d) Mass drainage rate versus prominence mass, displaying a positive linear correlation.
 } \label{mass-rates.eps}
 \end{figure}
%

%--- spatial integration for mass flux ---
After obtaining the downflow velocity $v(s_i, t)$ at all cuts,	%($i=0,1,...23$), 
next we need the knowledge of the spatial width and temporal duration of each downflow stream.
For the former, we assumed that within each $6 \arcsec$ wide cut, 
no more than one vertical thread (typical width $2 \arcsec$) undergoes downflow at any moment,
  %\footnote{This assumption introduces an uncertainty up to a factor of $6 \arcsec/ 2 \arcsec= 3$.
  %Using narrower cuts would not help, because not every thread is exactly vertical and a narrower cut
  %tends to lose threads that it attempts to track.}
	%Namely, each downflow trajectory in the space-time diagram is caused by only one such thread,
giving an effective filing factor of $2 \arcsec/ 6 \arcsec=1/3$.
For the latter, we defined a switch function $Z(s_i, t)$ that describes whether downflow 
occurs ($Z=1$) or not ($Z=0$) at the $i$th cut and time $t$, 
depending on whether or not the emission intensity $I(s_i, t)|_{h=h_{\rm ref}}$ 
(e.g., Figure~\ref{downflow-promin.eps}(d)) in the corresponding space-time diagram
at height $h_{\rm ref}$ is $5\sigma$ above the pre-event background. 
%

%--- mass drainage rate result ---
Finally, summing over all vertical cuts (as spatial integration) 
yields the mass drainage rate:
 \beq
 \dot{M}_{\rm drain}(t) = \sum_{i=0}^{23} \rho A_{\rm thrd} v(s_i,t) Z(s_i, t),
 \label{Mdot_drain.eq} \eeq
where $A_{\rm thrd}=d_{\rm thrd}^2$ is the cross-sectional area of the thread. 
	%and the summation is taken over all vertical cuts as spatial integration.		%and takes into account spatial non-uniformity.
	%The resulting mass drainage rate is shown 
As shown in Figure~\ref{mass-rates.eps}(c), it has a mean  		% in red
 $$ % \beq 
 \langle \dot{M}_{\rm drain} \rangle = 1.1 \E{10} \g \ps ,		%4.2 \E{10}%1.4 \E{11} 
 $$ % \label{M_dot_drain_mean_eq} \eeq 
and a maximum
 $$ % \beq 
 \dot{M}_{\rm drain, \, max} = (2.1 \pm 0.5) \E{10} \g \ps		%(8.0 \pm 1.7) \E{10} % (2.7 \pm 0.6) \E{11} 
 $$ % \label{M_dot_drain_max_eq} \eeq 
at 11/26 14:08~UT.		%, which is $2.8$ times greater than the peak mass change rate $\dot{M}_{\rm max}$.
Integrating $\dot{M}_{\rm drain}$ over time yields the accumulative mass drained to date ${M}_{\rm drain}(t)$,
  % \beq
  % {M}_{\rm drain}(t) = \int_0^t \dot{M}_{\rm drain}(t) dt ,
  % \label{M_drain.eq} \eeq
as shown in Figure~\ref{mass-rates.eps}(b). At the end of the 19~hr duration, it reaches
 $$ % \beq 
 {M}_{\rm drain, \, max}= 7.7 \E{14} \g ,		%2.9 \E{15}	%9.7 \E{15} 
 $$ % \label{M_drain_max_eq} \eeq
which is $13$ times the maximum prominence mass $M_{\rm prom, \, max}$.

	%peak value of $2.9 \E{11} \g \ps$ at 11/26 13:03~UT that is $3.1$ times greater than the peak mass change rate. 
	%amounts to $M=1.2 \E{16} \g$ 

%----------------------------------------------------------
\subsection{Mass Condensation Rate $\dot{M}_{\rm cond}$}
\label{subsect_condense}

%--- M_cond result ------
The mass conservation equation~(\ref{mass.conserve.eq}) then gives the condensation rate 
 \beq
 \dot{M}_{\rm cond} = \dot{M}_{\rm prom} + \dot{M}_{\rm drain} ,
 \eeq
as shown in Figure~\ref{mass-rates.eps}(c).		%in green 
It closely follows the mass drainage rate $\dot{M}_{\rm drain}$, which dominates over the
prominence mass change $\dot{M}_{\rm prom}$ by about an order of magnitude.
	%, except during the first 1.5~hr of the initial condensation. 
The condensation rate has a similar mean
 $$ % \beq 
 \langle \dot{M}_{\rm cond} \rangle = 1.2 \E{10} \g \ps ,		%4.4 \E{10}%1.5 \E{11}
 $$ % \label{M_dot_cond_mean_eq} \eeq 
and maximum
 $$ % \beq 
 \dot{M}_{\rm cond, \, max} = (2.4 \pm 0.4) \E{10} \g \ps 		%(8.9 \pm 1.5) %(3.0 \pm 0.5) \E{11}
 $$ % \label{M_dot_cond_max_eq} \eeq 
at 11/26 11:37~UT.		%, concurrent with the maximum mass change rate $\dot{M}_{\rm max}$.
The cumulative mass condensed in 19~hr is
 $$ % \beq 
 {M}_{\rm cond, \, max}= 8.0 \E{14} \g ,		%3.0 \E{15}	%1.0 \E{16}
 $$ % \label{M_cond_max_eq} \eeq
about the same as the mass drained to date.

%--- mass drainage rate temporal variation, correlation with total mass ---
We find two episodes of rapid condensation and mass drainage,
starting at 11/26 07:00 and 22:00~UT, 	%with the second being about twice weaker.
	%These two episodes 
which coincide with the variations of the prominence mass (Figure~\ref{mass-rates.eps}(b)).
	%and spatial extent (Figures~\ref{mass-rates.eps}(c) and (b)).
In fact, the drainage rate is positively correlated with the total prominence mass,
with a linear correlation coefficient of $r=0.92 \pm 0.1$ (Figure~\ref{mass-rates.eps}(d)).
	%This is expected, because the more material the prominence has, the more mass it drains.
	%We will further discuss the implication of this correlation in Section~\ref{sect_discussion}.

%=======================================================================================
\section{Summary and Discussion}	%Summary and Discussion Conclusion
\label{sect_discussion}

We have presented new evidence of coronal condensation that leads to prominence formation
at magnetic dips after a confined eruption.
	%and coronal rain underneath a nearby dip.	% after an eruption, thanks to \sdo AIA's advanced imaging capabilities.
	%We summarize our findings and discuss 
Our major findings and their implications are as follows.

\begin{enumerate}	%====================

%---------------------------------------------------------------------------------
%\subsection{No Need for Equilibrium or Static Support}
%\label{subsect_no-support}

%---------------------------------------------------------------------------------
\item	%(1) %{\bf No Need for equilibrium or magnetostatic support.}

The estimated average mass condensation rate in the prominence 	%takes place at an average
is $\approx$$10^{10} \g \ps$, 		%and a maximum of $ (3.0 \pm 0.5) \E{11} \g \ps $. amounting to $9.7 \E{15} \g \approx 10^{16} \g $
giving an {\it accumulative} mass of $\sim$$10^{15} \g$ in 19~hr.
Most (96\%) of this mass, an order of magnitude more than
the {\it instantaneous} total mass of the prominence, is drained as vertical threads.
In other words, a {\it macroscopically quiescent} prominence is a 
{\it microscopically dynamic} object, with a total mass determined by the delicate balance between a 
significant mass supply and a drainage by downflows.		%slower than free-fall downflows.
	%object in delicate balance that involves significant continuous mass supply	%from condensation 
	%and mass drainage through downflows. 
This was suggested long ago
\citep[e.g.,][]{Priest.Smith.siphon-prominence.1979SoPh...64..267P, Tandberg-HanssenE.prominenceBook.1995nsp..book.....T},
but to the best of our knowledge, our observations provide the first quantitative evidence
in terms of the prominence mass budget.
%In this sense, hydrostatic support of the prominence material against gravity
%may not be necessary.

%Therefore, contrary to conventional prominence models, force balance or equilibrium support may not be necessary.

%---------------------------------------------------------------------------------
\item	%(1) linearly correlated

The mass condensation and drainage rates are very close 
and {\it linearly correlated} with the total prominence mass.	%with a correlation coefficient of $r=0.92 \pm 0.1$. 
Thus,	%This is intuitively expected, because 
the more material the prominence has, the more mass it is capable of draining.
Theoretical considerations of energy balance by	%Moreover, theoretical calculation by 
Low et al.~(2012b, in preparation) indicates that the total mass 
	%in a \citet{Kippenhahn.Schluter.promin-model.1957ZA.....43...36K} sheet
is a critical parameter that determines how fast condensation takes place.

%---------------------------------------------------------------------------------
\item	%(1) slower than free fall

The prominence downflows are much slower than free fall,
with an average acceleration of $\langle a_{\rm prom} \rangle \approx 1/6 /g_\Sun $,
similar to that found by \citet{ChaeJ.promin.threads.descend-knots.2010ApJ...714..618C}.
	% and even the coronal rain sliding down the nearby loops.
%This suggests that there is a relatively larger portion of gravity being
%canceled in the prominence,	% than in the coronal rain,
%possibly by the Lorentz force from the horizontal component of the magnetic field 
This shows that gravity in the prominence is countered significantly,
possibly by an upward Lorentz force
\citep[e.g.,][]{Low.Petrie.prominence.2005ApJ...626..551L, 
vanBallegooijen.Cramer.tangled-field-promin.2010ApJ...711..164V}
from the horizontal component of the magnetic field,
as indicated by polarimetric measurements \citep{CasiniR.prominence-horizontal-field.2003ApJ...598L..67C},
or by transverse wave generated pressure as originally suggested for braking coronal rain
\citep{Antolin.coronal-rain-Hinode-SOT.2011ApJ...736..121A}.
	%which may also be applicable to prominences because of their commonly
	%observed oscillations \citep{LinYong.promin-thread-review.2011SSRv..158..237L}.

%---------------------------------------------------------------------------------
%\subsection{Source of Material and Magnetic Environment}
%\label{subsect_B-config}

\item	%{\bf  Magnetic environment.}

After hours of cooling from $\sim$$10^{6} \K$ throughout		%to $\sim$$10^{5} \K$
almost the entire transequatorial loop system that confined the early eruption,
the initial prominence condensation ($\sim$$10^{5} \K$) occurs at the loop dips
  %\footnote{It was though found that condensation can occur without a dip 
  %\citep{Karpen.no-dip-promin.2001ApJ...553L..85K}.}
	%of these loops	% in a ``V" shape
and consequently progresses toward greater heights.		% (see Section~\ref{sect_cooling}). 
This suggests that these large-scale loops may function like a funnel with a vast 
collecting volume that feeds mass to the prominence,
and the dips likely provide favorable conditions for further cooling and condensation
\citep[cf.,][]{Karpen.no-dip-promin.2001ApJ...553L..85K}.
	% with cooled mass.
	%This suggests that these large-scale
	%loops function like a funnel with a vast collecting volume and feed the prominence with cooled mass.
	%that is consequently drained to the chromosphere. 

%---------------------------------------------------------------------------------

\item	% Confined eruption

The preceding eruption might be independent of the prominence reformation
or have provided heated material to make the confining transequatorial loops overdense, 
favoring radiative cooling 		% and condensation,
and possibly responsible for other post-eruption prominence formations
\citep{LinYong.cloud-promin.2006SPD....37.0121L, 
WangJX.ejecta-form-promin.transequator-loop.2007SoPh..244...75W}.

%---------------------------------------------------------------------------------

%\item	%{\bf  relation w/ eruption, source of material.}

%We propose that the passage of mass on the order of $10^{15} \g$/day
%could be a general property of prominences exhibiting active mass drainage,
	%regardless of the exact source of the mass.
	%separate from the issue of the mass source.
%including those long-lived, polar-crown prominences
%residing in coronal cavities 		%also exhibit active mass drainage
%likely fed by the ``magneto-thermal convection" suggested by \citet{Berger.hot-bubble.2011Natur.472..197B}.

%The formation of this prominence following the eruption, as also observed elsewhere	%in other events
%\citep[e.g.,][]{WangJX.ejecta-form-promin.transequator-loop.2007SoPh..244...75W},
%suggests that some of the condensed mass is originally brought up by the eruption into the transequatorial loops.
%Such post-eruption prominences, often seen as ``cloud prominences" 
%\citep{LinYong.cloud-promin.2006SPD....37.0121L}, may represent a distinct 
%sub-species from the polar crown quiescent prominences forming in coronal cavities,	% over large scale polarity reversal regions
%likely fed by the ``magneto-convection" suggested by \citet{Berger.hot-bubble.2011Natur.472..197B}. 
%The commonality is that any prominence, regardless of its origin, is a return flow of the mass cycle between the
%chromosphere and corona.

%---------------------------------------------------------------------------------
\item	%(1) {\bf Prominence mass and its implication for CME initiation, in light of Low (2001)}
		% See BC's email comment June 7.

% combine old #5 and #6 together

We propose that the passage of mass on the order of $10^{15} \g$/day, 
comparable to a CME mass, could be a general property of prominences exhibiting active drainage.
If the ``magneto-thermal convection" suggested by \citet{Berger.hot-bubble.2011Natur.472..197B}
in prominence-cavity structures also involves the drainage of such a large mass, 
its weight is energetically significant for holding a magnetic flux rope in the cavity until it erupts
into a CME \citep{LowBC.CME-review.2001JGR...10625141L, Zhang.Low.CME-nature.2005ARA&A..43..103Z}.

\end{enumerate}		%====================

%=======================================================================================
%\section{Conclusion}	%Summary and Discussion Conclusion
%\label{sect_conclude}

This study and future observations of coronal condensation	% with \hinodeA/SOT and \sdoA/AIA
may provide further insights into prominences as an MHD phenomenon, 
with answers to the long-standing questions noted in the Introduction.

%A more in-depth multiwavelength study of this event and comparison with theoretical models are underway and
%will be published in the future.

%In our view, this study heralds a new era of observations of coronal condensation made possible 
%with the new \hinodeA/SOT and \sdoA/AIA instruments.
%We expect to refine our analysis 	%and carry out simultaneous observations of the photospheric and coronal vector
	%magnetic field in the near future, which in particular, can help shed light on
	%in particular, 
%to investigate the responsible magnetic geometry of the prominence and loop dip
%and address other long-stand questions noted earlier.	% in regard to Items 5--7 above.

%=======================================================================================

\acknowledgments
{	%The \Ha images were provided by HAO MLSO. 
 WL was supported by AIA contract NNG04EA00C. 
 TEB was supported by the Solar-B FPP contract NNM07AA01C at Lockheed Martin.
 The National Center for Atmospheric Research is sponsored by the National Science Foundation.
 We thank the anonymous referee for critical comments and
 Sara Martin, Yong Lin, Jingxiu Wang, and Roberto Casini for helpful discussions.

	% in revision, at end of Section 2, after "to the formation of a new prominence", add
	% "which is likely a 304~\AA\ conterpart of so-called ``cloud prominences" seen in \Ha (S. Martin, private communication)."
}

%=======================================================================================
%\clearpage

% To-do at submission
% 1) To shorten long author lists: edit ms.bbl later by hand and simply do latex, not to invoke new bibtex entry.

{\scriptsize
%\bibliography{bib/ads_all_edit,bib/LiuW-group,bib/Liu-Wei}

}

%=======================================================================================

\end{document}